# Efficient, gigapixel-scale, aberration-free whole slide scanner using angular ptychographic imaging with closed-form solution

**Shi Zhao,[†] Haowen Zhou,[†] Siyu (Steven) Lin, Ruizhi Cao, and Changhuei Yang[*]**

*Department of Electrical Engineering, California Institute of Technology, Pasadena, California 91125, USA*
[†]*These authors contributed equally to this work*

*[*]chyang@caltech.edu*

**Abstract:** Whole slide imaging provides a wide field-of-view (FOV) across cross-sections of biopsy or surgery samples, significantly facilitating pathological analysis and clinical diagnosis. Such high-quality images that enable detailed visualization of cellular and tissue structures are essential for effective patient care and treatment planning. To obtain such high-quality images for pathology applications, there is a need for scanners with high spatial bandwidth products, free from aberrations, and without the requirement for z-scanning. Here we report a whole slide imaging system based on angular ptychographic imaging with a closed-form solution (WSI-APIC), which offers efficient, tens-of-gigapixels, large-FOV, aberration-free imaging. WSI-APIC utilizes oblique incoherent illumination for initial high-level segmentation, thereby bypassing unnecessary scanning of the background regions and enhancing image acquisition efficiency. A GPU-accelerated APIC algorithm analytically reconstructs phase images with effective digital aberration corrections and improved optical resolutions. Moreover, an auto-stitching technique based on scale-invariant feature transform ensures the seamless concatenation of whole slide phase images. In our experiment, WSI-APIC achieved an optical resolution of 772 nm using a 10×/0.25 NA objective lens and captures 80-gigapixel aberration-free phase images for a standard 76.2 mm × 25.4 mm microscopic slide.

## 1. Introduction

Whole slide imaging (WSI) digitizes tissue sections on microscope slides into high-resolution images that can be stored, analyzed, and shared, thereby advancing the shift from traditional pathology to digital diagnosis [1,2]. Recently, the advent of digital image analysis using deep learning has transformed digital pathology in numerous clinical and biological applications [3–5]. WSI plays a significant role in revealing both local and global spatial tissue and cell interactions [6,7], thereby providing crucial insights for clinical diagnosis.

These data-centric techniques heavily rely on the quality of digital images and the availability of a large amount of data [2,8]. Current WSI platforms face a significant technical challenge in meeting the growing demand for high-quality and high-resolution images within the digital pathology community. Aberrations, including defocus, can cause image blur or distortion, resulting in information loss or misrepresentations in downstream analysis [9,10]. Potential solutions include the use of z-scanning and autofocusing systems to address defocus aberrations and reduce information loss during imaging [11–14]. However, no autofocusing metric can be robustly generalized to all types of samples [15]. Additionally, other aberration terms of the optical system, such as field-dependent aberrations, cannot be corrected efficiently and conveniently.

Aside from these technical challenges, the use of deep learning for digital pathology analysis have also increase the demand for large WSI data sets, so that the deep learning systems can learn to read through the variations associated with individual slides. High quality and consistent WSI imaging technology can provide an alternative pathway to address this large data demand – high consistency in WSI images can cut down on extraneous variations and allow the deep learning model to focus and train on truly relevant pathological features with much smaller data sets. In the WSI pipeline, from sample preparations to image processing, the histological staining procedure introduces the most variation, leading to significant discrepancies in digital images [16]. To date, no staining protocol can effectively addresse these stain variations. A promising approach to overcome this issue is to eliminate the staining process from the pipeline and provide the structure and morphology of the samples with phase images. This is because cellular structures are generally associated with refractive index distributions [17] that are directly captured in phase images.

To achieve high-quality and high-resolution complex field imaging and address the aforementioned challenges, several advanced techniques can be considered, including digital holographic microscopy [18–20], transport-of-intensity [21–24], coherent diffractive imaging [25,26], and ptychographic imaging techniques [27–29]. These techniques are capable of providing amplitude-phase information, which reflects changes in absorption and refractive index in tissue sections. However, digital holographic microscopy necessitates a reference beam for interferometry and requires prior knowledge of the system's aberration for corrections [19]. Transport-of-intensity methods use multiple images near the focal plane along the axial direction to retrieve phase information. Unfortunately, the reliance on specific focal plane positions and boundary condition limits its applicability in current WSI systems. Coherent diffractive imaging recovers phase images from several axial diffraction patterns obtained from a coherent source; however, it suffers from a reduction in optical resolution by a factor of two compared to conventional brightfield microscopy.

Ptychographic imaging holds promise for digital pathology applications due to its ability to perform high-resolution imaging by taking measurements with lateral overlaps and employing numerical optimizations to compensate for aberrations and retrieve phase information. A compact and efficient ptychographic imaging modality has been proposed and developed with phase imaging and aberration correction capabilities [30] Another promising technique is Fourier ptychographic microscopy (FPM) [31–34]. FPM utilizes computational methods to achieve high spatial-bandwidth-product imaging. It collects a series of low-resolution images under oblique illuminations and uses an iterative algorithm to reconstruct high-resolution images while maintaining a large field-of-view (FOV) with a small numerical aperture (NA) objective lens. This technique can also determine and numerically correct aberrations, thereby directly addressing defocus and system aberrations computationally [35]. However, both ptychographic imaging and FPM are prone to failure under severe system aberrations, which often occur at the peripheral regions of each FOV in WSI [36] Moreover, the iterative reconstruction algorithm in FPM suffers from challenges related to parameter tuning due to its nonconvex optimization nature, making it less robust for large-scale automated reconstructions in distinctive pathological applications.

Angular ptychographic imaging with a closed-form method (APIC) addresses robustness concerns in current ptychographic imaging reconstruction [36]. APIC utilizes the Kramers–Kronig relations for complex field reconstruction [37–39] and employs analytical techniques to correct aberrations while extending high spatial frequency content through darkfield measurements. Using NA-matching and darkfield measurements, APIC achieves high-resolution, aberration-free complex field retrieval with a low-magnification and large-FOV objective. This approach demonstrates exceptional robustness against aberration correction.

In this study, we introduced WSI-APIC (whole slide imaging based on APIC), a whole slide scanner that provides efficient, large-FOV, aberration-free, complex field imaging at the gigapixel-scale. This study aimed to adapt the implementation of WSI-APIC for digital pathology WSI applications through engineering design optimizations and algorithmic accelerations. The following sections detail our system pipeline and the numerical algorithms employed. Section 3 presents the validation and evaluation of our technique through experiments and analyses. Finally, we summarize the performance of our system and discuss potential improvements for future high-throughput digital pathology applications.

## 2. Methods

A general pipeline of the WSI-APIC system is illustrated in Fig. 1. The system operates in three steps. (i) A sample-locating system employs oblique incoherent illumination from a light emitting diode (LED) lamp to perform initial high-level auto-segmentation of the sample area. This process generates a scanning mask that helps avoid unnecessary scanning of background regions. (ii) The WSI-APIC system scans the sample areas of the slide according to the predefined scanning mask. For each FOV, a series of images is captured under different illumination angles provided by LEDs. Reconstructions of these images is then performed using a GPU-accelerated APIC algorithm. (iii) An auto-stitching system automatically combines the reconstructed images to produce a whole slide image.

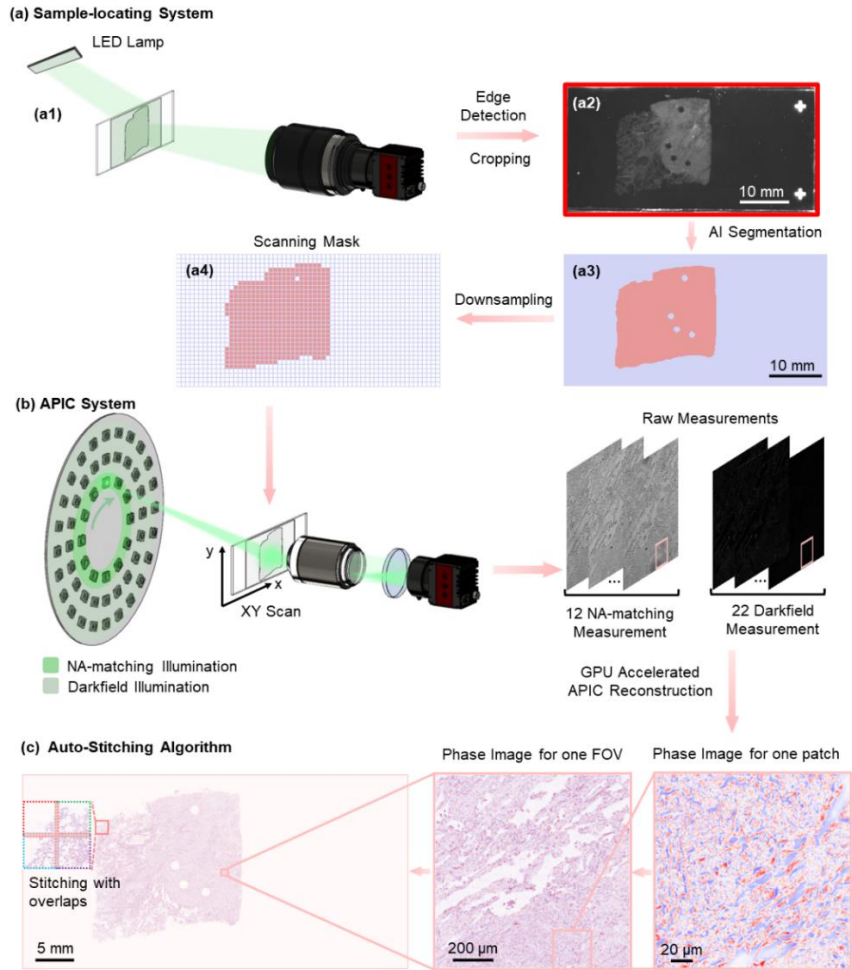

Fig. 1. General pipeline of the WSI-APIC system. (a) The sample-locating system auto-segments sample areas and generates a scanning mask for each slide. (b) The WSI-APIC microscope, equipped with customized LED illuminations, scans the sample according to the scanning mask and numerically reconstructs high-resolution, aberration-free images of the complex field. (c) An auto-stitching system integrates the reconstructed images to produce the whole slide images. NA stands for numerical aperture.

*2.1 Sample-Locating System*

In our study, the sample-locating system comprised an LED lamp, a sample holder, and a lens-camera imaging component. This configuration enabled the pre-identification of approximate sample locations, facilitating smart scanning during whole slide imaging. In this study, we implemented a simple lens imaging system with oblique incoherent illumination from the LED lamp, as shown in Fig. 1(a1). This setup generated a whole slide brightfield image, clearly depicting the profile of the sample. It can be adapted to both stained and unstained histological samples. For stained samples, the natural absorption of tissue areas provided image contrast to distinguish between the background glass slide and sample content. For unstained samples, the slight refractive index differences between tissue sections and the immersion medium resulted in scattered light under oblique illumination. This approach, similar to oblique illumination microscopy [40,41], provided image contrast that aided in differentiating the sample from the background.

For our study, an edge-detection algorithm based on peak identification was initially applied to identify the boundaries of the coverslip and slide. This was followed by cropping to eliminate regions outside the coverslip, ensuring that the segmentation was focused solely on the areas of interest within the slide (Fig. 1(a2)). Subsequently, the cropped image, which conformed to the boundaries of the coverslip, was processed using a pretrained segment anything model (SAM) [42] with zero-shot learning capabilities. Although SAM had not been specifically trained on microscopic slide images, it generated several moderately accurate masks. The imperfectness of these masks might be attributed to spatial intensity variations within tissue structures. To refine the precise sample mask, we further developed a mask selection algorithm. First, we filtered out masks with low accuracy ratings and those covering small segmented areas. Given the prior knowledge that intensity variations within the sample were relatively small compared to the sample-background interface, we calculated the mean and variance of intensities for all SAM-generated masks. Based on these metrics, we automatically identified and selected the masks corresponding to the sample region, merging them into a single mask representing the entire sample area.

The processed sample mask was prepared for the scanner. Based on the FOV of our camera, we calculated the numbers of scans required to cover the entire sample area in the lateral directions. Subsequently, the sample mask was downsampled to create a low-resolution scanning mask that aligns with the number of scanning positions needed (Fig. 1(a4)). This scanning mask avoided excessive scanning during microscopic imaging, thereby optimizing the time required for image acquisition and processing.

*2.2 WSI-APIC Setup and Forward Model*

In our study, the WSI-APIC facilitated the analytical reconstruction of complex fields and enabled robust aberration estimation and correction with a simple setup and minimal modifications to a conventional brightfield microscope. As illustrated in Fig. 1(b), a programmable LED disk was positioned in front of the sample to provide quasi-plane-wave illumination from various oblique angles. The sample was then imaged using a standard 4$f$ microscopy system including an objective lens, a tube lens, and a camera. The WSI-APIC system employed two types of LEDs illumination. First, LEDs with illumination angles matching the NA of the objective lens were sequentially activated to produce NA-matching measurements. Second, LEDs with illumination angles exceeding the objective lens's receiving angle were lit for darkfield measurements, which required a longer exposure time. All the measurements were subsequently used for reconstruction.

To effectively demonstrate the imaging system, we modeled the forward process from illumination to image acquisition. A 2D thin sample was illuminated by a plane wave emitted by the $i^{th}$ LED ($i = 1, 2, ..., n$), corresponding to a transverse k-vector $\boldsymbol{k}_i$. The modulated sample spectrum $\hat{S}_i$ at the camera plane is given by

$$\hat{S}_i(\boldsymbol{k}) = \hat{O}(\boldsymbol{k}-\boldsymbol{k}_i)H(\boldsymbol{k}) = \hat{O}(\boldsymbol{k}-\boldsymbol{k}_i)\text{Circ}_{\text{NA}}(\boldsymbol{k})e^{j\phi(\boldsymbol{k})}, \qquad (1)$$

where $\hat{O}$ represents the sample's spectrum, $\boldsymbol{k}$ denotes the transverse spatial frequency coordinates, and $H$ is the coherent transfer function (CTF) of the imaging system. The CTF is a circular function $\text{Circ}_{\text{NA}}$ with an NA-dependent radius and a phase term $\phi$ depicting system aberrations. Due to the Fourier transform property, changing the illumination angles shifts the CTF laterally, sampling different regions of the sample's spectrum. Finally, the $i^{th}$ intensity measurement $I_i(\boldsymbol{x})$ obtained from the camera is,

$$I_i(\boldsymbol{x}) = \left|\left[\mathcal{F}^{-1}(\hat{S}_i)\right](\boldsymbol{x})\right|^2 = |S_i(\boldsymbol{x})|^2. \qquad (2)$$

where $\mathcal{F}^{-1}$ denotes the inverse Fourier transform operator.

### 2.3 WSI-APIC reconstruction

The graphical schematic of the WSI-APIC reconstruction algorithm is shown in Fig. 2. The algorithm comprises two main steps: brightfield reconstruction using Kramers–Kronig relations [37–39], incorporating aberration correction for NA-matching measurements and spectrum extension using darkfield measurements. Aberrations in the system were estimated during the brightfield reconstruction phase and corrected for both NA-matching and darkfield spectra.

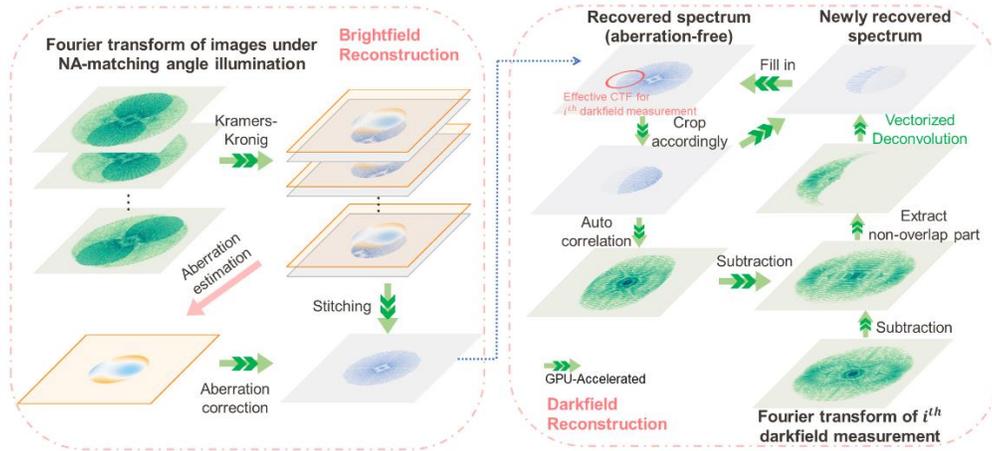

Fig. 2. Reconstruction pipeline of the WSI-APIC system. It contains two major steps: brightfield reconstruction and darkfield reconstruction. In the brightfield reconstruction process, the Kramers–Kronig relation was employed to recover the corresponding spectra. Then, phase differences between overlapping spectra were then utilized to identify and correct the aberrations of the imaging system. These reconstructed spectra were then corrected for aberrations and stitched together. During the darkfield reconstruction part, the known spectrum from the $i^{th}$ darkfield measurement was used to isolate the cross-correlation component from other autocorrelation terms. By solving a linear equation involving the isolated cross-correlation, the unknown spectrum was analytically determined. We applied GPU-acceleration to original WSI-APIC reconstruction algorithm. The green arrows indicate GPU-accelerated processes.

Based on the forward model of the WSI-APIC system, the spectrum of the sample under a given NA-matching illumination angle at the camera plane was constrained by a circular function support, with the zero-frequency (DC) component located at the edge of this support. To facilitate the application of the directional Hilbert transform in brightfield reconstruction, it was advantageous to shift the DC component to the center of the spatial frequency plane. This shift simplified the reconstruction process.

According to the properties of the Fourier transform, shifting by $-\boldsymbol{k}_i$ in Fourier space is equivalent to multiplying by a phase shift of $e^{-2\pi k_i \cdot x}$ in the spatial domain. We introduced an important relation between the sample's spectrum at the camera plane $\hat{S}_i(\boldsymbol{k})$ and the shifted spectrum $\hat{S}'_i(\boldsymbol{k}) = \hat{S}_i(\boldsymbol{k} + \boldsymbol{k}_i)$:

$$\left|\mathcal{F}^{-1}\left(\hat{S}'_i(\boldsymbol{k})\right)\right|^2 = \left|\mathcal{F}^{-1}\left(\hat{S}_i(\boldsymbol{k}+\boldsymbol{k}_i)\right)\right|^2 = \left|S_i(\boldsymbol{x})e^{-2\pi k_i \cdot x}\right|^2 = |S_i(\boldsymbol{x})|^2 = I_i(\boldsymbol{x}). \quad (3)$$

The intensity measurement $I_i(\boldsymbol{x})$ is invariant to the addition of phase ramp. Hence, reconstructing $S'_i(\boldsymbol{x}) = \mathcal{F}^{-1}\left(\hat{S}'_i(\boldsymbol{k})\right)$ or $S_i(\boldsymbol{x}) = \mathcal{F}^{-1}\left(\hat{S}_i(\boldsymbol{k})\right)$ is equivalent when using $I_i(\boldsymbol{x})$.

The first step in the WSI-APIC reconstruction process involved applying the Kramers–Kronig relations, which analytically linked the real and imaginary parts of a complex function. These relations could be used to reconstruct the complex field using measurements obtained under illumination angles that exactly match the maximal acceptance angle of the objective lens (NA-matching measurements). We constructed an auxiliary function by taking the logarithm of the shifted complex field $S'_i(\boldsymbol{x})$ in a point-wise manner and adding a constant phase term:

$$\log\left[S'_i(\boldsymbol{x})e^{-j\theta_i}\right] = \log[|S'_i(\boldsymbol{x})|] + j\arg[S'_i(\boldsymbol{x}) - \theta_i], \quad (4)$$

where $\theta_i = \phi(\boldsymbol{k}_i)$ is a constant phase offset defined by the pupil phase at $\boldsymbol{k} = \boldsymbol{k}_i$. As the intensity $I_i(\boldsymbol{x})$ is measured, the real part of the auxiliary function is simple to obtain. The imaginary part is computed using the directional Hilbert transform in conjunction with the intensity measurements.

$$G(\boldsymbol{k}) \coloneqq \left[\mathcal{F}\left(log\left[S'_i(\boldsymbol{x})e^{-j\theta_i}\right]\right)\right](\boldsymbol{k}) = \begin{cases} [\mathcal{F}(\log I_i)](\boldsymbol{k}), & \boldsymbol{k} \cdot \boldsymbol{k}_i < 0 \\ 0.5[\mathcal{F}(\log I_i)](\boldsymbol{k}), & \boldsymbol{k} \cdot \boldsymbol{k}_i = 0 \\ 0, & \boldsymbol{k} \cdot \boldsymbol{k}_i > 0 \end{cases} \quad (5)$$

where $\mathcal{F}$ denotes the Fourier transform operator. Then, the desired field with a constant phase offset can be restored using the inverse Fourier transform and exponential function.

$$S'_i(\boldsymbol{x})e^{-j\theta_i} = \exp\left[\mathcal{F}^{-1}\left(G(\boldsymbol{k})\right)\right]. \quad (6)$$

The reconstructed complex field from the WSI-APIC system includes aberration terms. To handle these aberrations, we worked in the spatial frequency domain and utilized the phases of multiple reconstructed spectra. The spectrum of the constructed complex field is given by:

$$\hat{S}'_i(\boldsymbol{k})e^{-j\theta_i} = \mathcal{F}\left(S'_i(\boldsymbol{x})e^{-j\theta_i}\right) = \hat{A}(\boldsymbol{k})e^{j\hat{\alpha}(\boldsymbol{k})}Circ_{NA}(\boldsymbol{k}+\boldsymbol{k}_i)e^{j\phi(\boldsymbol{k}+\boldsymbol{k}_i)-j\theta_i}, \quad (7)$$

where $\hat{A}(\boldsymbol{k})$ represents the amplitude of the sample's spectrum and $\hat{\alpha}(\boldsymbol{k})$ is the phase term. For two parts of samples' spectra captured under different LED illumination angles, we define

$$\mathcal{S}_{il} \coloneqq \{\boldsymbol{k}|Circ_{NA}(\boldsymbol{k}+\boldsymbol{k}_i)Circ_{NA}(\boldsymbol{k}+\boldsymbol{k}_l) \neq 0\}. \quad (8)$$

$\mathcal{S}_{il} \neq \emptyset$ when the two parts of the spectra has overlap regions. Within the overlapped region $\mathcal{S}_{il}$, we calculated the phase difference between the two spectra:

$$\arg(\hat{S}_i'(\boldsymbol{k})e^{-j\theta_i}) - \arg(\hat{S}_l'(\boldsymbol{k})e^{-j\theta_l}) = [\hat{\alpha}(\boldsymbol{k}) + \phi(\boldsymbol{k}+\boldsymbol{k}_i) - \theta_i] - [\hat{\alpha}(\boldsymbol{k}) + \phi(\boldsymbol{k}+\boldsymbol{k}_l) - \theta_l]$$
$$= [\phi(\boldsymbol{k}+\boldsymbol{k}_i) - \phi(\boldsymbol{k}+\boldsymbol{k}_l)] - [\theta_i - \theta_l]. \tag{9}$$

When we considered the phase difference of the two spectra, the contribution from the sample spectrum canceled out, and the difference depended solely on the aberration of the system. The remaining phase difference was linear with respect to the aberration functions. By analyzing all overlapping regions (suppose $m$ in total), we formulated a linear problem by applying a linear operator:

$$D\phi = \beta_\Delta, \tag{10}$$

where $D := \begin{bmatrix} D_{i_1 l_1} - D^0_{i_1 l_1} \\ D_{i_2 l_2} - D^0_{i_2 l_2} \\ \ldots \\ D_{i_m l_m} - D^0_{i_m l_m} \end{bmatrix}$,

$$[(D_{il} - D^0_{ij})\phi][\boldsymbol{K}] = \phi(\boldsymbol{K}+\boldsymbol{K}_i) - \phi(\boldsymbol{K}+\boldsymbol{K}_l) - \phi(\boldsymbol{K}_i) + \phi(\boldsymbol{K}_l), \quad \boldsymbol{K} \in \mathcal{S}_{il},$$

$$\beta_\Delta := \begin{bmatrix} \arg\left(\hat{S}'_{i_1}(\boldsymbol{K})e^{-j\phi(\boldsymbol{K}_{i_1})}\right) - \arg\left(\hat{S}'_{l_1}(\boldsymbol{K})e^{-j\phi(\boldsymbol{K}_{l_1})}\right), & \boldsymbol{K} \in \mathcal{S}_{i_1 l_1} \\ \arg\left(\hat{S}'_{i_2}(\boldsymbol{K})e^{-j\phi(\boldsymbol{K}_{i_2})}\right) - \arg\left(\hat{S}'_{l_2}(\boldsymbol{K})e^{-j\phi(\boldsymbol{K}_{l_2})}\right), & \boldsymbol{K} \in \mathcal{S}_{i_2 l_2} \\ \ldots \\ \arg\left(\hat{S}'_{i_m}(\boldsymbol{K})e^{-j\phi(\boldsymbol{K}_{i_m})}\right) - \arg\left(\hat{S}'_{l_m}(\boldsymbol{K})e^{-j\phi(\boldsymbol{K}_{l_m})}\right), & \boldsymbol{K} \in \mathcal{S}_{i_m l_m} \end{bmatrix}.$$

Here, $\boldsymbol{K}$ represents the flattened vector of 2D spatial frequency $\boldsymbol{k}$, $\boldsymbol{K}_n$ is the flattened position of illumination vector $\boldsymbol{k}_n$, and $\mathcal{S}_{i_n l_n}$ denotes the $n^{th}$ overlapping region. Upon solving this equation, the aberrations of the imaging system were extracted, enabling the correction of the initially reconstructed spectra. These corrected spectra were subsequently assembled to generate an aberration-free image with a twofold enhancement in resolution.

The corrected spectra served as prior knowledge for the darkfield reconstruction process (as shown in the lower part of Fig. 2). This process involved using darkfield measurements to extend the spatial frequency spectrum of the sample.

By selecting an appropriate NA for illumination, a relatively high degree of overlapping was obtained among the spectra of the NA-matching and darkfield measurements. The Fourier transform of each darkfield measurement comprised cross-correlation between the known and unknown spectra, as well as their autocorrelations. We defined $\hat{U}_i$ as the unknown spectrum and $\hat{P}_i$ as the known spectrum. The relationship can be expressed as follows:

$$I_i(\boldsymbol{x}) = \left|\mathcal{F}^{-1}(\hat{U}_i + \hat{P}_i)\right|^2 = |U_i(\boldsymbol{x}) + P_i(\boldsymbol{x})|^2$$
$$= U_i(\boldsymbol{x})U_i^*(\boldsymbol{x}) + U_i(\boldsymbol{x})P_i^*(\boldsymbol{x}) + P_i(\boldsymbol{x})U_i^*(\boldsymbol{x}) + P_i(\boldsymbol{x})P_i^*(\boldsymbol{x}). \tag{11}$$

The Fourier transform of $I_i$ is given by

$$[\mathcal{F}(I_i)](\boldsymbol{k}) = \hat{U}_i \star \hat{U}_i + \hat{U}_i \star \hat{P}_i + \hat{P}_i \star \hat{U}_i + \hat{P}_i \star \hat{P}_i. \tag{12}$$

where $\star$ denotes correlation operator. As $\hat{P}_i \star \hat{P}_i$ was already known, we focused on the cross-correlation parts and the autocorrelation of the unknown spectrum. These three terms had different supports, allowing us to isolate the nonoverlapping components. In regions $\mathcal{Q}_i$ where the cross-correlation terms do not overlap with the autocorrelation of the unknown spectrum, we obtain a equation with respect to the unknown component:

$$[\mathcal{F}(I_i)](\boldsymbol{k}) - [\hat{P}_i \star \hat{P}_i](\boldsymbol{k}) = [\hat{P}_i \star \hat{U}_i](\boldsymbol{k}), \quad \boldsymbol{k} \in \mathcal{Q}_i. \tag{13}$$

To solve for $\hat{U}_i$, we could apply a similar strategy used in deconvolution. By constructing a block Toeplitz matrix for a specific region, we could convert the correlation operation $\hat{P}_i \star \hat{U}_i$ into a matrix multiplication $C\hat{U}_i$, thereby transforming the problem into a set of linear equations. Notably, the matrix $C$ had dimensions $M \times N$, where $M$ represents the number of nonzero points in the linear region $\mathcal{Q}_i$ corresponding to the number of linear equations from Eq. (13)), and $N$ denotes the number of nonzero points in the region of the unknown spectrum. This set of linear equations can be solved to extend the retrieved spectrum across all darkfield measurements. Finally, a high-resolution, aberration-free complex field could be reconstructed.

## *2.4 Algorithm Acceleration*

In WSI, each slide typically encompasses thousands of FOVs containing sample information. Often, the reconstruction of high-quality image data from this massive volume is a significant processing bottleneck. The original APIC algorithm, designed for execution on central processing units (CPUs), offers sufficient speed for many applications. However, in the context of WSI-APIC, the unprecedented volume of data necessitates further acceleration of the algorithm.

To address this challenge, we adapted the original APIC algorithm for execution on a PyTorch-based graphic processing unit (GPU) platform [43]. A general workflow is illustrated in Fig. 2, with green arrows indicating the GPU computation workflow. Experimental testing and evaluation indicated that, with the exception of the aberration estimation algorithm, all other computational operations could be efficiently adapted for GPU processing through appropriate coding. It is important to note that the primary bottleneck speed in the existing algorithm is related to darkfield reconstruction, specifically in the construction of the convolution matrix $C$ used to solve $[\hat{P}_i \star \hat{U}_i](\boldsymbol{k})$ in Eq. (13). The construction of matrix $C$ involves the following two steps: (i) forming multiple Toeplitz matrices from the rows of the known spectrum $\hat{P}_i$; and (ii) assembling the convolution matrix $C$ from these Toeplitz matrices, incorporating the nonzero constraints of the unknown region $\hat{U}_i$ and the nonoverlapping region $\mathcal{Q}_i$. This process accounts for approximately 90% of the total reconstruction time.

Given that optimizing this step/function was crucial for accelerating the algorithm, we redesigned it using a vectorized algorithm. We noticed that to fill in the convolution matrix $C$, the original APIC algorithm employed nested loops to filter indices corresponding to columns in the nonoverlapping region $\mathcal{Q}_i$ and the unknown region $\hat{U}_i$, identifying nonzero elements in $\hat{U}_i$. This approach, which relies on extensive usage of loops and positional condition checks, was time-consuming. In our GPU implementations, we adopted a vectorized construction method to suit parallel computing capabilities. We calculated all indices from the unknown region $\hat{U}_i$ corresponding to a given column in the nonoverlapping region $\mathcal{Q}_i$ simultaneously. This approach allowed us to only loop over the nonoverlapping region $\mathcal{Q}_i$ and filtered out multiple rows of $C$ at once, rather than handling them one small block at a time. To illustrate, constructing the convolution matrix $C$ could be simplified as setting up linear equations, where the nonoverlapping region limits us to $M$ equations and the unknown region limits us to $N$ unknowns. The original APIC algorithm loops over both the unknowns and the equations, determining coefficients for each specific equation and unknown. In contrast, our APIC-GPU algorithm employed a vectorized approach to compute all coefficients for each equation concurrently, significantly accelerating the process. This optimization resulted in a 45% reduction in the time required for this step, leading to a 40% decrease in overall reconstruction time in our experiments.

Despite the algorithmic design, we observed a significant increase in processing speed for the Kramers–Kronig and darkfield reconstructions when implemented on GPU. The parallel computing capabilities of GPUs notably enhanced the performance of large matrix operations. The sole exception to this improvement was the aberration estimation step. The construction of the difference operator $D$, as described in Eq. (10), required extensive index search operations and numerous small, complex matrix computations. Our analysis indicates that these operations performed more efficiently on CPUs compared to GPUs. This discrepancy was primarily due to PyTorch's limited support and optimization for very small, complex matrix operations, in contrast to libraries such as numpy [44] on CPUs. Furthermore, in our application, GPUs were less efficient at handling the irregular memory access patterns associated with advanced indexing operations compared to CPUs. Consequently, we opted to perform the construction of $D$ on the CPU.

*2.5 Auto-stitching Algorithm*

In practice, the position accuracy of translational stages is never perfect. Even precise translational stages can exhibit micro-level misalignments. Therefore, a robust and efficient stitching algorithm capable of compensating for misalignments is crucial within the entire pipeline. To address this challenge in our study, overlapping regions between adjacent scans were retained during the scanning process. With the aid of these overlapping regions and the scanning mask, we implemented an auto-stitching algorithm based on scale-invariant feature transform (SIFT) [45]. The algorithm began its processing with the top-left corner of the slide. If a FOV was identified as the background in the scanning mask, the corresponding area remained blank. Otherwise, overlapping regions between this FOV and its adjacent previously stitched sample regions were cropped. The SIFT algorithm was employed to detect distinctive key points within these overlapping regions. These key points were then matched across the overlapping regions of the previously stitched sample areas based on their feature descriptors. Using the matched key points, we estimated the geometric transformation required to align this FOV with the previously stitched sample regions. Finally, the overlapping regions were blended to smoothen any visible seams. In cases where an FOV was entirely surrounded by background, its placement was determined based on its position in the scanning mask.

### 3. Results

In our experiments, we used a 10×/0.25 NA objective lens (Plan N, Olympus) coupled with a camera (Allied Vision Prosilica GT6400), mounted on a microscope body (Olympus IX51) for data acquisition. The camera featured a pixel pitch of 3.45 µm, with each FOV comprising 2560 × 2560 pixels, corresponding to a physical size of 957 µm × 957 µm. Given that the standard dimensions of pathology slides were 76.2 mm × 25.4 mm × 1 mm, a 60 × 30 grid scanning mask was employed, with a 6% overlap rate, to accommodate for x-y scanner instability and errors (typically 5 µm for our Thorlabs MLS203-1 XY scanning stage). For illumination, we employed an RGB LED disk (DotStar RGB LED Disk, Adafruit), controlled by an Arduino board (Arduino Uno, Arduino). During the acquisition, 12 LEDs were sequentially illuminated in a ring configuration for NA-matching illuminations, while 22 LEDs in two outer rings were used for darkfield illuminations, achieving an illumination NA of 0.44. Unless otherwise specified, a green LED with a central wavelength of 523 nm was employed for the experiments. Whole slide scanning was performed using a motorized x-y scanning stage (MLS203-1, Thorlabs) to achieve a high-precision control of sample positioning.

*3.1 Siemens Star for System Quantification*

To quantify the resolution achieved by our system, we imaged a Siemens star target. We compared the results obtained from a brightfield microscope using 10×/0.25 NA and 20×/0.4 NA objective lens with images reconstructed by APIC-CPU and APIC-GPU. The images are shown in Figs. 3(a-d), respectively. To quantitatively compare the optical resolution of these images, we identified the smallest radius at which 40 peaks along the line were preserved, known as the sparrow limit [29]. The circles at this radius are highlighted in Figs. 3(a-d). The radial profiles along these circles are plotted in Fig. 3(e). We observed that the captured image using a brightfield microscope with a 10×/0.25 NA objective lens was unable to resolve any stripes. However, the images reconstructed using APIC-CPU and APIC-GPU could significantly distinguish these stripes, yielding results similar to those obtained from a brightfield microscope with a 20x 0.4 NA objective lens, which was considered the ground truth. The radial profiles of the original APIC-CPU and APIC-GPU in Fig. 3(e) exactly overlapped. Additionally, for the whole Siemens star FOV, the mean squared error between APIC-CPU and APIC-GPU was on the order of $10^{-7}$, which originated from floating point precision differences and was negligible.

We also determined that the resolution was approximately 772 nm. The illumination NA corresponding to the maximum illumination angle was approximately 0.44, while the objective NA was 0.25. In our experiments, we used green LEDs with a central wavelength of 522 nm, resulting in a theoretical resolution of $\lambda/(NA + NA_{illum}) = \frac{522\ nm}{0.44+0.25} = 761$ nm. This validation demonstrated that our system's optical resolution was very close to the theoretical value.

Our APIC-GPU significantly decreased the reconstruction time while maintaining identical results and resolution. Figure 3(f) illustrates a runtime comparison between the original APIC-CPU algorithm and our APIC-GPU algorithm as a function of image patch size. In the comparison experiments, the reconstructed sample was a Siemens star target with varying patch sizes. The original APIC-CPU algorithm was run on an Intel Core i7-14700KF CPU, while the APIC-GPU was run on an NVIDIA GeForce RTX 4090 GPU. As depicted in Fig. 3(f), for all patch sizes, the reconstruction time with APIC-GPU was significantly shorter than that of the original APIC-CPU algorithm, with the difference becoming more pronounced as the patch size increased. Specifically, when the patch side length was 512 pixels, the original APIC-CPU algorithm took approximately 500 s to complete the reconstruction, whereas APIC-GPU required only 10.8 s. We achieved a speed boost up to 50 times under our demonstrated settings. This rapid reconstruction speed made APIC feasible for application in WSI. Additionally, the significantly reduced reconstruction time allowed us to use larger patch sizes (e.g., 512 pixels) for reconstruction, which could mitigate the lateral shifts effects at large defocus distances [46] and helped reduce stitching artifacts to some extent.

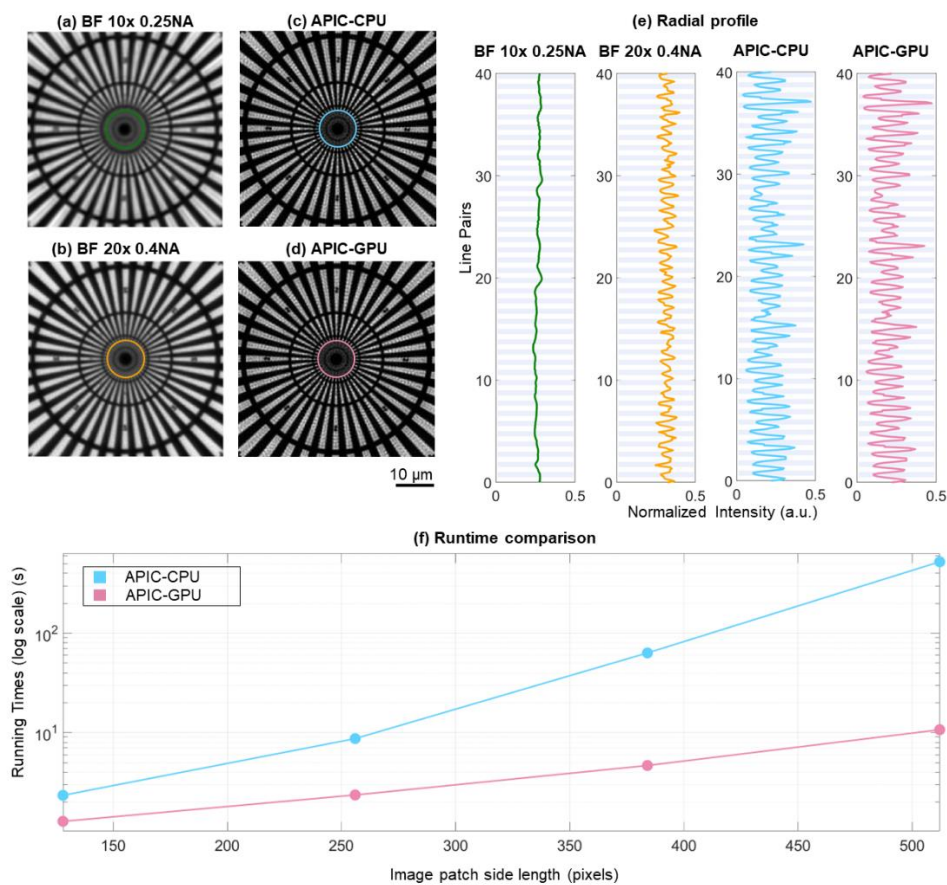

Fig. 3. Comparative analysis of resolution quantification and reconstruction times. (a) Captured image under a brightfield microscope with a 10×/0.25 NA objective lens. (b) Captured Siemens star image under a brightfield microscope with a 20×/0.4 NA objective (set as ground truth). (c) Reconstructed Siemens star intensity image using APIC-CPU. (d) Reconstructed Siemens star intensity image using APIC-GPU. (e) Radial profiles along the circles highlighted in (a)–(d). The circles represent the smallest radius where 40 peaks along the line were preserved for reconstructed amplitude of Siemens star using APIC-CPU and APIC-GPU, indicating a resolution of 772 nm. Notably, APIC-CPU and APIC-GPU had the same radial profile, with the two lines exactly overlapped. (f) Runtime comparison between APIC-CPU and APIC-GPU. The patch sizes used in this simulation formed an arithmetic sequence ranging from 128 to 512 pixels, with a common difference of 128 pixels.

## 3.2 Whole Slide Imaging for Unstained Slides

As previously noted, phase imaging presents a promising solution to the issue of stain variation by delivering high-contrast phase information without additional staining. This phase information, such as optical path length differences in tissue sections, can uncover the micron-scale organization and support pathological assessments [47,48]. To validate our methods, we first performed phase imaging on an unstained early-stage non-small-cell lung cancer pathology slide. Given that this unstained sample was nearly transparent, conventional brightfield imaging lacked contrast and failed to provide useful tissue structural information. Instead, we utilized green LED illumination and retrieved phase information using WSI-APIC. Prior to scanning, we positioned the system's FOV approximately at the sample center of and adjusted the objective to bring the sample into rough focus, making no further adjustments during the scanning process. For reconstruction, each 2560 × 2560 pixel FOV from the camera was divided into 25 patches of 512 × 512 pixels, which were individually reconstructed and then stitched together. Our auto-stitching algorithm then combined all FOVs according to the scanning mask. The resulting phase image, depicted in Fig. 4(a), measured 429,940 × 216,280 pixels with an optical resolution of 778 nm, providing a vast amount of information. The phase images and their corresponding reconstructed pupil functions for different regions of the sample, including various cells and tissues, are shown in Fig. 4(b1)–(b6). The phase images clearly outlined the physiological structures of red blood cells, tumor cells, lymphocytes, and other cellular components, assisting pathologists in evaluating cancerous conditions. The reconstructed pupil functions in Fig. 4(b4) and (b6) highlight significant aberrations near the sample center, possibly due to sample thickness or uneven slide placement causing defocus. Unlike traditional whole slide scanners, which require autofocusing and other complex methods, our system employed the APIC reconstruction algorithm to directly correct defocus-induced aberrations, resulting in high-quality, aberration-free phase reconstruction. The depth of field attainable by our WSI-APIC system is further detailed in the Supplementary Material.

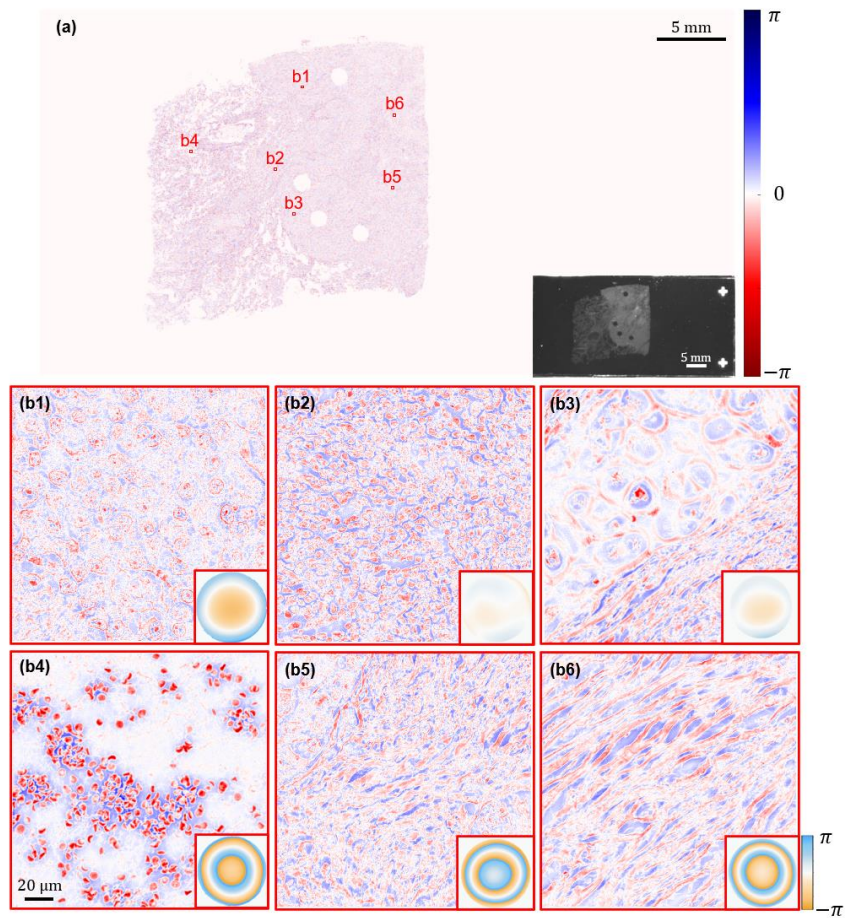

Fig. 4. (a) Whole slide phase image of an unstained non-small-cell lung cancer sample. (b) Zoomed-in views of phase images from different positions

For the inset slide image shown in Fig. 4, our sample-location system identified samples in only 525 out of the 1800 FOVs covering the entire slide. Given that the average time to capture one FOV of WSI-APIC raw data was 34 s, the total acquisition time was 4.97 h. This indicates a significant improvement, reducing the acquisition time by 70.5% compared to a naïve full scan. This efficiency enhancement significantly accelerated the data acquisition process. Typically, this selective scanning approach can reduce image collection time by approximately 20%–80%, depending on the density and distribution of the sample on the slide.

Regarding the reconstruction time, our APIC-GPU algorithm reconstructed one patch in 10.7 s (25 patches in one FOV). Notably, to further avoid unnecessary reconstructions, we implemented an intensity variance threshold for images patches. If the low-resolution image patches exhibited an intensity variance below this threshold, they were classified as background and skipped during reconstruction. This strategy significantly enhanced our reconstruction efficiency, resulting in a total slide reconstruction time of 33.3 hours. In contrast, naively applying the APIC-CPU method would have taken an estimated 1925 hours, approximately 58 times longer than our optimized method. Notably, our current reconstruction time is constrained by available computational resources. By utilizing parallelize processing across multiple GPU kernels on computing units, it will significantly reduce reconstruction time by a factor of 10 to 20. Detailed discussions on related speed improvements can be found in Section 4.

Compared to traditional APIC, our WSI-APIC system significantly reduced both scanning and reconstruction times, making whole slide phase imaging practically achievable. While our approach is currently less time-efficient compared to conventional whole slide scanners, it offers an effective and easy way for obtaining aberration-free, high-quality, high-resolution whole slide phase images. This method is especially suitable for pathology applications that require high-quality, staining-free sample information, such as deep learning analysis with limited data settings. The speed performance of our WSI-APIC system could be further enhanced by incorporating high power LEDs, hardware triggers, optimized illumination design, and parallel processing using multiple GPUs. These potential improvements will be discussed in detail in the Conclusions and Discussion Section.

*3.3 Whole Slide Imaging for Stained Slides*

Here, we report results of a whole slide scan of a hematoxylin and eosin (H&E)-stained human tuberculosis slide (Carolina Biological Supply Inc.). For this process, we executed the entire WSI-APIC pipeline on each of the three RGB channels separately. We extracted the amplitude from the final reconstructed complex field for each channel, combined the amplitudes, and performed white balancing. Figure 5(a) displays the reconstructed whole slide color image. Figs. 5(b1)–(b3) present some zoomed-in patches with their corresponding reconstructed pupil aberrations, while Figs. 5(c1) –(c3) shows their corresponding brightfield microscope images with a 10×/0.25 NA objective lens obtained during the scanning without focus adjustment. Figs. 5(d1)–(d3) illustrates corresponding brightfield microscope images captured using Olympus BX51 microscope with a 20×/0.4 NA objective lens. They were well-focused and designated as the ground truth. The color difference arises from the varying post-processing and white balancing strategies employed by different microscopes. Due to chromatic aberrations, brightfield microscope images often exhibit noticeable blur and artifacts, especially when defocus is significant (Fig. 5(c3)). However, in our study, WSI-APIC effectively corrected both inherent system chromatic aberrations and those caused by defocus, resulting in high-quality H&E scans. It's clear from the images (comparing Fig. 5(b1) –(b3), to the corresponding Fig. 5(c1) –(c3) images) that WSI-APIC was able to provide improved image quality.

Although our scanner took longer for capture and reconstruction compared to conventional scanner, our approach effectively mitigated imaging aberrations, yielding higher image quality. Moreover, our scan eliminated the need for autofocus and other complex procedures, making the scanning system simpler and more cost effective.

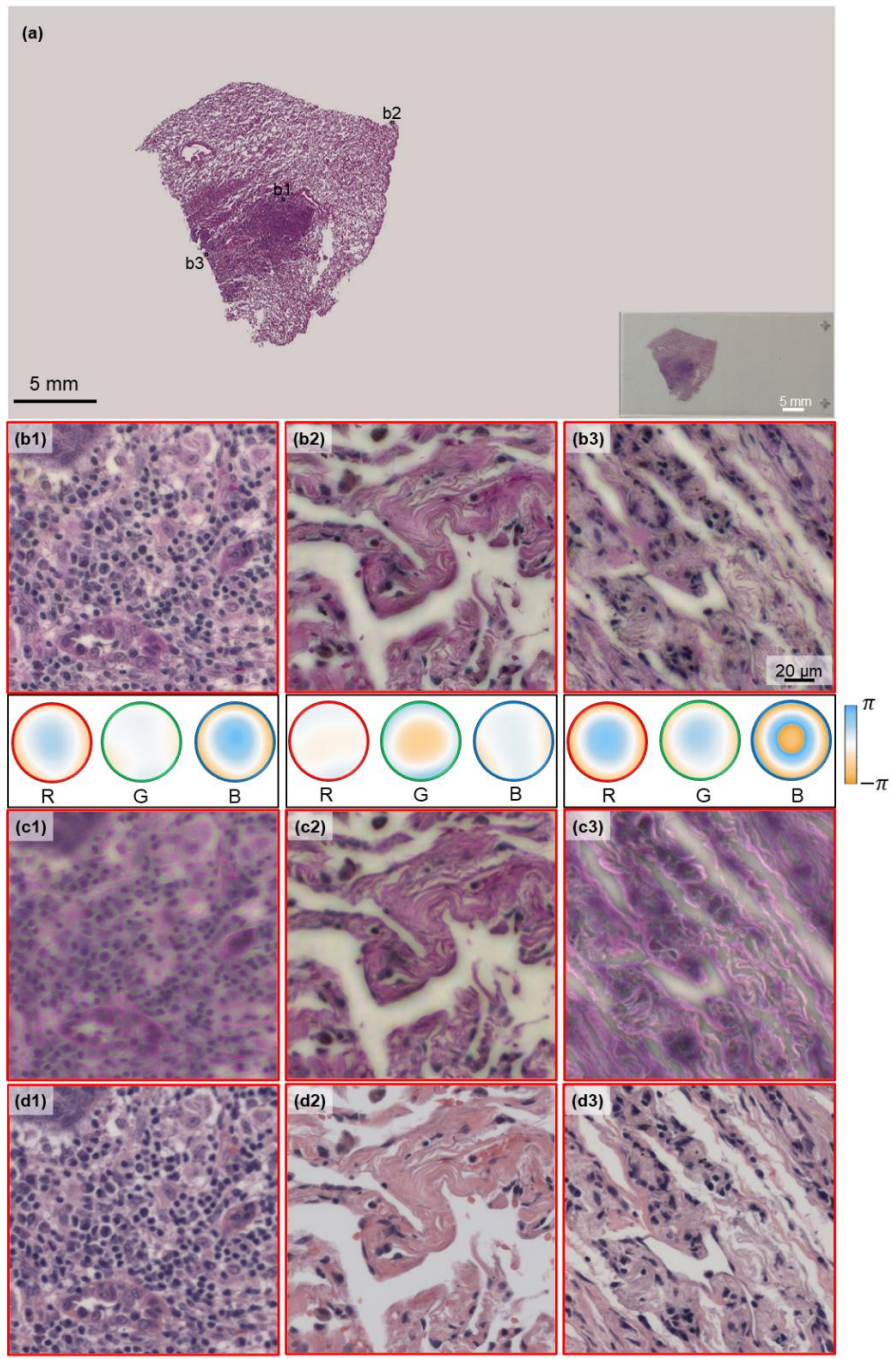

Fig. 5. (a) Whole slide image of an H&E-stained non-small-cell lung cancer sample acquired with the WSI-APIC system. (b) Zoomed-in views of images from distinct positions in (a) and their corresponding reconstructed pupil aberrations for the three color channels, respectively. R: Red, G: Green, B: Blue. (c) Captured images of corresponding views under a brightfield microscope with a 10×/0.25 NA objective lens when scanning. (d) Captured focused images of corresponding views under Olympus BX51 commercial brightfield microscope with a 20×/0.4 NA objective lens (set as ground truth).

## 4. Conclusions and Discussion

We reported the successful implementation and demonstration of WSI-APIC - an adaptation of analytical computational microscopy for digital pathology applications. We believe that this technology is a potentially viable solution to address the growing demand for high-quality and high-consistency data in digital pathology deep learning analyses. Building on the APIC framework, we designed a sample-locating system and incorporated a GPU-accelerated APIC algorithm. This approach reduced the acquisition time by up to 80% and accelerated the reconstruction speed by up to 50 times, demonstrating the practical feasibility of implementing APIC in WSI. Our WSI-APIC system achieved an optical resolution of 745 nm using a 10×/0.25 NA objective lens and customized LED illuminations. We also demonstrated the capability to produce 80-gigapixel aberration-free WSI for both stained and unstained samples. The pipeline proposed in this study offers a promising prototype for advancing the development of high-throughput, z-scan free, high-quality scanners.

Despite significant advancements made with the WSI-APIC technique for WSI application, our current WSI-APIC system remains relatively slow and has several areas where further advancements could be pursued. Future studies could focus on a few directions to build a commercial-grade whole slide scanner based on WSI-APIC.

First, the current acquisition speed is primarily limited by both the power of the LEDs and synchronization issues among the system components. The LEDs used in our system currently deliver a typical power of 6 mW for the green channel, with a 120-degree divergence angle and a LED-to-sample distance of 103.4 mm. This setup results in an illumination power at the sample region in the microwatt range. Consequently, the acquisition times are approximately 0.1 s per frame for NA-matching illumination and 0.5 s per frame for darkfield illumination. These time expenses are further extended by imperfect synchronization among the camera, translation stage, and LED disk, resulting in a total acquisition time of 34 s per FOV. Upgrading to high power LEDs with a customized printed circuit board design or employing laser sources with galvo mirror could increase illumination brightness by 50–100 times compared to the current LEDs. Implementing hardware triggers can improve synchronization between the translational stages, LEDs, and camera. With these modifications, we can potentially reduce the acquisition time by 50–100 times compared to the current setup. Such a setup has the potential to enable a 76.2 mm × 25.4 mm whole slide scan to be completed in just a few minutes.

Second, our current setup uses 12 NA-matching illuminations and 22 darkfield illuminations to achieve a synthetic NA of 0.69. There is potential to reduce redundancy in this setup. According to the original APIC work [36], the proportion of the unknown spectrum within the captured spectrum should not exceed 42% to ensure effective reconstruction in darkfield measurements. Our current illumination setup, however, only utilizes up to 17% of this potential. Thus, by redesigning the LED positions, we can expect to achieve the same resolution with fewer measurements in each FOV. This adjustment would reduce both acquisition and reconstruction times linearly. Employing multiplexing strategies, such as polarization multiplexing or color multiplexing, could further decrease acquisition times. However, these strategies may trade off some spatial-bandwidth-product.

In the context of algorithm improvements, a simple but effective way is to upgrade the computing units. By utilizing server-level computing units, it is possible to parallelize the processing of different patches across multiple GPU kernels. This upgrade could potentially reduce reconstruction time by a factor of 10 to 20 compared to current GPU capabilities, particularly through the use of an Nvidia A100 computing node.

Building on the fundamental work of the WSI-APIC system, the aforementioned potential improvements can be pursued in future developments. Collectively, our objective is to enhance the efficiency and feasibility of the WSI-APIC system, with the ultimate aim of developing a commercial-grade WSI-APIC capable of delivering high-quality and high-consistency WSI for high-throughput pathology applications.

**Funding.** Heritage Research Institute for the Advancement of Medicine and Science at Caltech (Grant No. HMRI-15-09-01), Rothenberg Innovation Initiative (RI2) in conjunction with the Hagopian Innovation Prize (Grant No. 25570017)

**Acknowledgments.** We would like to thank Dr. Mark Watson and Dr. Richard Cote at School of Medicine, Washington University of St. Louis for providing non-small-cell lung cancer samples.

**Disclosures.** The authors declare no conflicts of interest.

**Data availability.** The code and example data will be available on GitHub and CaltechData upon the acceptance of the manuscript.

**Supplemental document.** See Supplement document for supporting content.

**References**

1. F. Ghaznavi, A. Evans, A. Madabhushi, and M. Feldman, "Digital Imaging in Pathology: Whole-Slide Imaging and Beyond," Annu. Rev. Pathol. Mech. Dis. **8**, 331–359 (2013).

2. L. Pantanowitz, N. Farahani, and A. Parwani, "Whole slide imaging in pathology: advantages, limitations, and emerging perspectives," PLMI 23 (2015).

3. S. Deng, X. Zhang, W. Yan, E. I.-C. Chang, Y. Fan, M. Lai, and Y. Xu, "Deep learning in digital pathology image analysis: a survey," Front. Med. **14**, 470–487 (2020).

4. A. Janowczyk and A. Madabhushi, "Deep learning for digital pathology image analysis: A comprehensive tutorial with selected use cases," Journal of Pathology Informatics **7**, 29 (2016).

5. M. Salvi, U. R. Acharya, F. Molinari, and K. M. Meiburger, "The impact of pre- and post-image processing techniques on deep learning frameworks: A comprehensive review for digital pathology image analysis," Computers in Biology and Medicine **128**, 104129 (2021).

6. M. Y. Lu, B. Chen, D. F. K. Williamson, R. J. Chen, M. Zhao, A. K. Chow, K. Ikemura, A. Kim, D. Pouli, A. Patel, A. Soliman, C. Chen, T. Ding, J. J. Wang, G. Gerber, I. Liang, L. P. Le, A. V. Parwani, L. L. Weishaupt, and F. Mahmood, "A Multimodal Generative AI Copilot for Human Pathology," Nature (2024).

7. H. Zhou, M. Watson, C. T. Bernadt, S. (Siyu) Lin, C. Lin, J. H. Ritter, A. Wein, S. Mahler, S. Rawal, R. Govindan, C. Yang, and R. J. Cote, "AI -guided histopathology predicts brain metastasis in lung cancer patients," The Journal of Pathology **263**, 89–98 (2024).

8. N. M. Atallah, M. S. Toss, C. Verrill, M. Salto-Tellez, D. Snead, and E. A. Rakha, "Potential quality pitfalls of digitalized whole slide image of breast pathology in routine practice," Modern Pathology **35**, 903–910 (2022).

9. J. R. Gilbertson, J. Ho, L. Anthony, D. M. Jukic, Y. Yagi, and A. V. Parwani, "Primary histologic diagnosis using automated whole slide imaging: a validation study," BMC Clin Pathol **6**, 4 (2006).

10. C. Massone, H. Peter Soyer, G. P. Lozzi, A. Di Stefani, B. Leinweber, G. Gabler, M. Asgari, R. Boldrini, L. Bugatti, V. Canzonieri, G. Ferrara, K. Kodama, D. Mehregan, F. Rongioletti, S. A. Janjua,

V. Mashayekhi, I. Vassilaki, B. Zelger, B. Žgavec, L. Cerroni, and H. Kerl, "Feasibility and diagnostic agreement in teledermatopathology using a virtual slide system," Human Pathology **38**, 546–554 (2007).

11.     R. Redondo, G. Bueno, J. C. Valdiviezo, R. Nava, G. Cristóbal, O. Déniz, M. García-Rojo, J. Salido, M. D. M. Fernández, J. Vidal, and B. Escalante-Ramírez, "Autofocus evaluation for brightfield microscopy pathology," J. Biomed. Opt. **17**, 036008 (2012).

12.     Y. Sun, S. Duthaler, and B. J. Nelson, "Autofocusing in computer microscopy: Selecting the optimal focus algorithm," Microscopy Res & Technique **65**, 139–149 (2004).

13.     A. Patel, U. G. J. Balis, J. Cheng, Z. Li, G. Lujan, D. S. McClintock, L. Pantanowitz, and A. Parwani, "Contemporary Whole Slide Imaging Devices and Their Applications within the Modern Pathology Department: A Selected Hardware Review," Journal of Pathology Informatics **12**, 50 (2021).

14.     L. Van Der Graaff, G. J. L. H. Van Leenders, F. Boyaval, and S. Stallinga, "Computational imaging modalities for multi-focal whole-slide imaging systems," Appl. Opt. **59**, 5967 (2020).

15.     Z. Bian, C. Guo, S. Jiang, J. Zhu, R. Wang, P. Song, Z. Zhang, K. Hoshino, and G. Zheng, "Autofocusing technologies for whole slide imaging and automated microscopy," Journal of Biophotonics **13**, e202000227 (2020).

16.     C. Dunn, D. Brettle, M. Cockroft, E. Keating, C. Revie, and D. Treanor, "Quantitative assessment of H&E staining for pathology: development and clinical evaluation of a novel system," Diagn Pathol **19**, 42 (2024).

17.     S. Yin, R. Cao, M. Liang, C. Shen, H. Zhou, O. Zhang, and C. Yang, "Can deep neural networks work with amplitude and phase input of defocused images?," Opt. Express **32**, 25036 (2024).

18.     F. Charrière, A. Marian, F. Montfort, J. Kuehn, T. Colomb, E. Cuche, P. Marquet, and C. Depeursinge, "Cell refractive index tomography by digital holographic microscopy," Opt. Lett. **31**, 178 (2006).

19.     M. K. Kim, "Principles and techniques of digital holographic microscopy," J. Photon. Energy 018005 (2010).

20.     B. Kemper and G. Von Bally, "Digital holographic microscopy for live cell applications and technical inspection," Appl. Opt. **47**, A52 (2008).

21.     M. Beleggia, M. A. Schofield, V. V. Volkov, and Y. Zhu, "On the transport of intensity technique for phase retrieval," Ultramicroscopy **102**, 37–49 (2004).

22.     L. Waller, L. Tian, and G. Barbastathis, "Transport of Intensity imaging with higher order derivatives," Opt. Express **18**, 12552 (2010).

23.     C. Zuo, Q. Chen, L. Tian, L. Waller, and A. Asundi, "Transport of intensity phase retrieval and computational imaging for partially coherent fields: The phase space perspective," Optics and Lasers in Engineering **71**, 20–32 (2015).

24.     H. Zhou, E. Stoykova, M. Hussain, and P. P. Banerjee, "Performance analysis of phase retrieval using transport of intensity with digital holography [Invited]," Appl. Opt. **60**, A73 (2021).

25.     G. J. Williams, H. M. Quiney, B. B. Dhal, C. Q. Tran, K. A. Nugent, A. G. Peele, D. Paterson, and M. D. De Jonge, "Fresnel Coherent Diffractive Imaging," Phys. Rev. Lett. **97**, 025506 (2006).

26. B. Abbey, K. A. Nugent, G. J. Williams, J. N. Clark, A. G. Peele, M. A. Pfeifer, M. De Jonge, and I. McNulty, "Keyhole coherent diffractive imaging," Nature Phys **4**, 394–398 (2008).

27. T. Wang, S. Jiang, P. Song, R. Wang, L. Yang, T. Zhang, and G. Zheng, "Optical ptychography for biomedical imaging: recent progress and future directions [Invited]," Biomed. Opt. Express **14**, 489 (2023).

28. S. Jiang, T. Wang, and G. Zheng, "Coded Ptychographic Imaging," in *Coded Optical Imaging*, J. Liang, ed. (Springer International Publishing, 2024), pp. 181–203.

29. L. Yang, R. Wang, Q. Zhao, P. Song, S. Jiang, T. Wang, X. Shao, C. Guo, R. Pandey, and G. Zheng, "Lensless polarimetric coded ptychography for high-resolution, high-throughput gigapixel birefringence imaging on a chip," Photon. Res. **11**, 2242 (2023).

30. C. Guo, Z. Bian, S. Jiang, M. Murphy, J. Zhu, R. Wang, P. Song, X. Shao, Y. Zhang, and G. Zheng, "OpenWSI: a low-cost, high-throughput whole slide imaging system via single-frame autofocusing and open-source hardware," Opt. Lett. **45**, 260 (2020).

31. G. Zheng, R. Horstmeyer, and C. Yang, "Wide-field, high-resolution Fourier ptychographic microscopy," Nature Photon **7**, 739–745 (2013).

32. G. Zheng, C. Shen, S. Jiang, P. Song, and C. Yang, "Concept, implementations and applications of Fourier ptychography," Nat Rev Phys **3**, 207–223 (2021).

33. X. Ou, R. Horstmeyer, C. Yang, and G. Zheng, "Quantitative phase imaging via Fourier ptychographic microscopy," Opt. Lett. **38**, 4845 (2013).

34. P. C. Konda, L. Loetgering, K. C. Zhou, S. Xu, A. R. Harvey, and R. Horstmeyer, "Fourier ptychography: current applications and future promises," Opt. Express **28**, 9603 (2020).

35. X. Ou, G. Zheng, and C. Yang, "Embedded pupil function recovery for Fourier ptychographic microscopy," Opt. Express **22**, 4960 (2014).

36. R. Cao, C. Shen, and C. Yang, "High-resolution, large field-of-view label-free imaging via aberration-corrected, closed-form complex field reconstruction," Nat Commun **15**, 4713 (2024).

37. Y. Baek and Y. Park, "Intensity-based holographic imaging via space-domain Kramers–Kronig relations," Nat. Photonics **15**, 354–360 (2021).

38. Y. Baek, K. Lee, S. Shin, and Y. Park, "Kramers–Kronig holographic imaging for high-space-bandwidth product," Optica **6**, 45 (2019).

39. C. Shen, M. Liang, A. Pan, and C. Yang, "Non-iterative complex wave-field reconstruction based on Kramers–Kronig relations," Photon. Res. **9**, 1003 (2021).

40. C. Sanchez, G. Cristóbal, G. Bueno, S. Blanco, M. Borrego-Ramos, A. Olenici, A. Pedraza, and J. Ruiz-Santaquiteria, "Oblique illumination in microscopy: A quantitative evaluation," Micron **105**, 47–54 (2018).

41. R. Sugimoto, R. Maruyama, Y. Tamada, H. Arimoto, and W. Watanabe, "Contrast enhancement by oblique illumination microscopy with an LED array," Optik **183**, 92–98 (2019).

42. A. Kirillov, E. Mintun, N. Ravi, H. Mao, C. Rolland, L. Gustafson, T. Xiao, S. Whitehead, A. C. Berg, W.-Y. Lo, P. Dollár, and R. Girshick, "Segment Anything," (2023).

43. A. Paszke, S. Gross, F. Massa, A. Lerer, J. Bradbury, G. Chanan, T. Killeen, Z. Lin, N. Gimelshein, L. Antiga, A. Desmaison, A. Köpf, E. Yang, Z. DeVito, M. Raison, A. Tejani, S. Chilamkurthy, B. Steiner, L. Fang, J. Bai, and S. Chintala, "PyTorch: An Imperative Style, High-Performance Deep Learning Library," (2019).

44. C. R. Harris, K. J. Millman, S. J. Van Der Walt, R. Gommers, P. Virtanen, D. Cournapeau, E. Wieser, J. Taylor, S. Berg, N. J. Smith, R. Kern, M. Picus, S. Hoyer, M. H. Van Kerkwijk, M. Brett, A. Haldane, J. F. Del Río, M. Wiebe, P. Peterson, P. Gérard-Marchant, K. Sheppard, T. Reddy, W. Weckesser, H. Abbasi, C. Gohlke, and T. E. Oliphant, "Array programming with NumPy," Nature **585**, 357–362 (2020).

45. D. G. Lowe, "Object recognition from local scale-invariant features," in *Proceedings of the Seventh IEEE International Conference on Computer Vision* (IEEE, 1999), pp. 1150–1157 vol.2.

46. H. Zhou, B. Y. Feng, H. Guo, S. (Steven) Lin, M. Liang, C. A. Metzler, and C. Yang, "Fourier ptychographic microscopy image stack reconstruction using implicit neural representations," Optica **10**, 1679 (2023).

47. T. L. Nguyen, S. Pradeep, R. L. Judson-Torres, J. Reed, M. A. Teitell, and T. A. Zangle, "Quantitative Phase Imaging: Recent Advances and Expanding Potential in Biomedicine," ACS Nano **16**, 11516–11544 (2022).

48. Z. Wang, G. Popescu, K. V. Tangella, and A. Balla, "Tissue refractive index as marker of disease," J. Biomed. Opt. **16**, 1 (2011).